# Logistic Knowledge Tracing: A Constrained Framework for Learner Modeling

Philip I. Pavlik, Jr., Luke G. Eglington, and Leigh M. Harrell-Williams

*Abstract*—Adaptive learning technology solutions often use a learner model to trace learning and make pedagogical decisions. The present research introduces a formalized methodology for specifying learner models, Logistic Knowledge Tracing (LKT), that consolidates many extant learner modeling methods. The strength of LKT is the specification of a symbolic notation system for alternative logistic regression models that is powerful enough to specify many extant models in the literature and many new models. To demonstrate the generality of LKT, we fit 12 models, some variants of well-known models and some newly devised, to 6 learning technology datasets. The results indicated that no single learner model was best in all cases, further justifying a broad approach that considers multiple learner model features and the learning context. The models presented here avoid student-level fixed parameters to increase generalizability. We also introduce features to stand in for these intercepts. We argue that to be maximally applicable, a learner model needs to adapt to student differences, rather than needing to be pre-parameterized with the level of each student's ability.

## I. Creating and Using Logistic Regression Learner Models to Inform Learning Technology Pedagogy

Logistic regression is a statistical method that has been used by many investigators to characterize student performance for various learning tasks. In this paper, we explain a formalized approach to creating logistic regression models that subsumes other methods and provides flexibility that allows better determination of an accurate model than off-the-shelf approaches like the Additive Factors Model (AFM), Instructional Factors Analysis (IFA), Performance Factors Analysis (PFA), PFA-Decay or Recent-PFA (R-PFA) [1]–[7]. We focus this work on building models that generalize to new learners for which there is no data, which is a typical situation when attempting to optimize adaptive instruction in a learning system.

This work is motivated by growing research interest to understand student learning in adaptive, personalized educational software. Adaptive personalization typically requires a model of student learning that captures the complexities of learning, but it can be unclear which of the many candidate models is ideal in a particular context. This paper introduces a theoretically motivated framework for systematic specification and evaluation of student models for adaptive instructional environments, Logistic Knowledge Tracing (LKT). The LKT framework also facilitates comparing the relative strengths and weaknesses of different learner models, enabling more productive research. The LKT framework promotes a better understanding of the possible models that could be specified given the factors at play in any learning system. Further, by delineating the possibilities for predictive features in a logistic regression model, the framework makes it easier for modelers to create and test new features.

The necessity of a flexible model development approach became clear over our work on the LearnSphere/DataShop project. The LearnSphere/DataShop project aims to facilitate sharing educational data and analyses (as of January 2021, Carnegie Mellon University DataShop contained more than 3400 datasets comprising 1 million hours of student learning data in which they learn a variety of content) [8]–[10]. Via LearnSphere, researchers develop and share components that allow others to run various models of learning that track student performance across time. The variety of datasets and educational contexts in the Datashop datasets made it clear that a great variety of logistic regression learner models can be specified for different contexts. However, until now, there has no principled framework for learner model development. Because logistic regression modeling is quite flexible, it leads to the question: What do these models have in common? By identifying the common elements of these logistic regression models (described below), it becomes possible to provide a unified system (an LKT component) in the LearnSphere for creating logistic regression learning models. The LearnSphere

Submitted August 8, 2019. This work was partially supported by the National Science Foundation Learnsphere (NSF #1443068) and the Learner Data Institute (NSF #1934745) projects and a grant from the Institute of Education Sciences (ED #R305A190448). Any opinions, findings, and conclusions or recommendations expressed in this material are those of the authors and do not necessarily reflect the views of IES or NSF. We wish to thank the High-Performance Computer Center at University of Memphis, which provided computer time for the analyses. (*Corresponding author*: Philip I Pavlik, Jr.)

P. I. Pavlik, Jr., is with the Institute of Intelligent Systems and Department of Psychology, University of Memphis, Memphis, TN, 38152 USA (email: ppavlik@memphis.edu).

L. G. Eglington is with the Institute for Intelligent Systems, University of Memphis, Memphis, TN, 38152 USA (email: lgglngtn@memphis.edu).

L. M. Harrell-Williams is with the Department of Counseling, Educational Psychology and Research and affiliate faculty with the Institute of Intelligent Systems, University of Memphis, Memphis, TN, 38152 USA (e-mail: leigh.williams@memphis.edu).

4LKT component implements the LKT framework, a unified system that permits easier creation and comparison of logistic regression models of correctness. We have also released LKT as an easy-to-use R package.

*A. The Utility of the Approach*

The primary utility of a model of student learning is to make inferences about learning that guide pedagogical decisions, i.e., decisions about what and when to teach. Such pedagogical decision making is the basis for any automated instructional system [11], [12]. Using the sequence of prior performances for each student practicing items (e.g., a specific test question), the model represents the likelihood each subsequent item may be correctly answered as a probability. Typically, this probability guides pedagogical decision making. For example, a low probability prediction implies that instruction is needed and that the item might be challenging for the student. In contrast, a high probability may indicate that instruction is less critical. In fact, a high probability prediction by the model implies that the item could be marked as learned since the probability is intended to be tracking learning. This method of using the probability is known as mastery learning, the pedagogical principle that skills should be practiced until learning reaches a pre-specified criterion (e.g., 95% correct prediction [13]).

However, a model is needed for many different pedagogical approaches. One relatively generic method (applicable to any model of performance) is shown by research that has suggested that selecting items for practice causes maximal learning at some fixed correctness probability (e.g., practice whichever item is closest to 80% correctness probability) [14], [15]. For example, Pavlik and Anderson [15] attempted to find this optimal practice threshold with a complex dynamic optimization that computed the probability correct at which learning was more efficient, given specific retention goals. This work found that a constant level of difficulty was optimal [26]. This conclusion about the presence of an optimal difficulty level has been supported by other methodological approaches based on an exhaustive search for an optimal schedule [14], [16]. The optimal difficulty will depend on features of the to-be-learned content (e.g. [17]).

Another use of this work is in creating open student models [18], in which the student can view the predictions the model has made about them. This feedback allows students to metacognitively monitor their performance more accurately, leading to more efficient learning if students address the deficiencies predicted by the model.

*B. History*

To better understand the goal of producing an accurate model using logistic regression for learner modeling, it is useful to survey the history of such an approach. The model we have created can be traced from the development of item response theory (IRT) [19], [20]. IRT was developed to understand and evaluate results from academic testing by allowing a modeler to fit parameters to characterize performance on a set of items by a set of students. Items typically refer to specific test or quiz items, but we extend the concept of an item to any single step of a task that is learned, including any immediate review following the task. Thus, the concept of an item is generalized to include practice items with or without subsequent feedback or review. As a specialized form of logistic regression, IRT predicts 0 or 1 (dichotomous) results, such as the results of individual items for students. In the case of a 1 item parameter IRT model, this result is predicted as a function of student ability minus item difficulty (x), which is scaled to a probability estimate from the logistic function cumulative distribution where $p = (1/(1+e^{-x}))$.

IRT has had a long developmental history, including the Linear Logistic Test Model (LLTM), which introduced the idea that multiple discrete "cognitive operations" may combine to determine an items' difficulty [21]. While traditional 1, 2, and 3 parameter IRT models might be called a behavioral description, their parameters do not map onto mental constructs. Thus, LLTM goes further by describing how latent traits combine to produce an item's difficulty. This work maps closely "rule spaces" or what have come to be known as Q-matrices, which are a means to specify the skills, concepts, knowledge, or cognitive operations needed to correctly answer a practice or test item [22], [23]. Many researchers use the term knowledge component (or KC) to refer generically to any of these learned proficiencies needed to respond correctly to practice items (e.g., [24]). For example, knowledge of how to apply the least common denominator to do fraction addition could be a KC.

However, the LLTM model alone does not capture learning. On a path toward modeling learning, we trace the development to work by Scheiblechner [25], which used the LLTM model, but also examined changes in difficulty as people repeated knowledge components. This work is well-reviewed by Spada and McGaw [26], who unpacked the history of these sorts of models, which have come to be known more recently as the additive factors model (AFM [7]) within the Artificial Intelligence in Education (AIED) and Educational Data Mining (EDM) communities. AFM is relatively straightforward and proposes that we add a term to these IRT-type models that capture each knowledge component's prior quantity of practice as a linear effect. These skill tracking models have been combined with the Q-matrix knowledge component models, perhaps because Tatsuoka's [22] work clearly explained this method from a less mathematical and more pedagogical perspective that has appealed to learning science researchers. A variant of this LLTM/AFM model is built into the DataShop repository at Carnegie Mellon University and is currently being used to do an automated search for better Q-matrixes to represent the skills in different educational systems [9].

Performance factors analysis (PFA) further improved the AFM model by fitting separate parameters for prior successes and failures to predict future performance. This change made logistic regression about as accurate as Bayesian knowledge tracing (BKT), a Markov model frequently used to fit similar data [6], [13], [27]. The PFA model assumes that there are two fundamental categories of practice, success and failure. As the psychological literature suggests, successes (in contrast to review after failing) may lead to more production-based learning and/or less forgetting [28]–[30] and may lead to automaticity [31]–[34]. In contrast, failure reveals a need for declarative learning of problem structure and verbal rules for production compilation [35]. We might expect learning after failure to depend most notably on the feedback that occurs after the task is complete. PFA methodology of categorizing event types within a KC was generalized into Instructional Factors Analysis (IFA), which explains that all event types may be tracked individually [36]. In addition to the possibility of using more terms in a logistic regression model, which IFA shows, we also might consider terms that are nonlinear. The simplest examples are terms that use the natural log function of the prior trials [36], [37]. The idea of nonlinear features opens a realm of complex possibilities, as shown in recent research. The earliest example of a nonlinear feature in the EDM/AIED literature appears to be the introduction of the PFA-Decay model [4], which elaborates on PFA by proposing that the counts of success and failures be exponentially weighted as a function of their recency in the stream of observations of performance for a KC. This weighting means that trials that occurred further in the past will have less effect. In terms of PFA, this means that if a student transitions from failure to success across a sequence of trials, the model can adjust its predictions quickly since past failures may have minimal effect relative to a history of recent successes. This weighting creates a greater sensitivity to recent changes in learning.

R-PFA is a variant of PFA-decay, which introduced another nonlinear component, a recency-weighted proportion of prior correct responses for a KC [3]. Since this feature is a proportion, the denominator must always be positive to compute the R-PFA feature. The model was created with the assumption of 3 failure "ghost" attempts to resolve this issue, making the proportion start at 0 and determining how sensitive the model is to further practice since these ghost attempts are used in computing the proportion. Like PFA-Decay, the attempts used to compute the proportion decay exponentially, so this feature is highly sensitive to performance changes. In later sections, we used a modified recency weighted proportion that had 1 ghost success and 1 ghost failure. These initial settings allow the model performance to trend downwards as well as upwards. We also apply the decaying proportion feature more generally because we found that it also works well to trace overall performance as a substitute for a student intercept parameter. In support of our goal for models that generalize to new data, the results of this paper document alternatives to using fixed or random student intercepts.

Concurrently with learner model development in psychometrics, EDM, and AIED communities, other fields are working on similar goals using logistic regression modeling. Memory researchers from psychological backgrounds have found strong evidence that memory follows consistent patterns of decay over time [38]–[42] and have researched how this decay may be reduced (e.g., [30], [41]). However, few learner models incorporate decay in their predictions. One recent example is the performance prediction equation (PPE), which has been under development for several years [38], [43]. This model bridges the gap between learner models and cognitive models because it more explicitly bases its structure on memory principles, including forgetting, spacing effects, and practice effects. PPE uses many features of the Adaptive Character of Thought - Rational (ACT-R) memory model [41]. While the PPE model has developed in the psychology cognitive modeling tradition, since it uses logistic regression (unlike ACT-R), it may be applicable for easy transfer to the psychometric, EDM, and AIED communities, if only it could be combined with other aspects of learner modeling, such as adaptivity and KC specific parameters. In this paper, we show how LKT allows the easy creation of such models.

We present two other models that account for forgetting, base2, and base4, which are also based on the ACT-R model. This suggests that there may be a family of ACT-R derivatives, and in this paper, we compare the three versions, PPE, base2, and base4. While the PPE equation uses all the ages since past practices to estimate forgetting, the base2 and base4 equations only use the first trial's age to estimate forgetting. In contrast, base 2 is much simpler, with one parameter scaling difference in interference and one parameter capturing forgetting. While base2 is not sensitive to spacing effect learning benefits in the data, both base4 and PPE use two parameters to scale the effects of practice according to the distribution (i.e., spacing) of that practice.

One may wonder why this paper does not include comparisons of other methods, such as BKT or DKT (Deep Knowledge Tracing). The main reason for this was to intentionally limit our scope to fully introduce the LKT methodology. This paper aims not to show which model is best or which modeling method is best, which others have done, most recently [44], but rather to open the world of logistic regression possibilities by providing a tool for researchers. This tool will enable richer comparisons with BKT or DKT if that is what the user wants. However, perhaps more meaningfully, the tool allows the user to tailor a model to the data characteristics in ways that are not currently possible with other methods. As we discuss, it is practical to create nonlinear features in addition to linear features by nesting the solution of these nonlinear features within a logistic model. This capability allows LKT models to use various parameterized nonlinear features that capture cognitive effects such as memory recency, the benefit of distributed practice, forgetting, and diminishing marginal returns for overlearning.

Further, using the logit from the logistic regression, it is possible to transform the logistic regression model predictions to predict future actions' latency. In addition to this flexibility, also argued for by others [3], one point we emphasize is the ease of use of the subsequent model in a running adaptive learning system as demonstrated in several ways, most recently including testing a system with an adaptive model of practice for learning Anatomy and Physiology textbook knowledge [15]–[17], [45], [46]. Unlike BKT, LKT does not typically require the full sequence of prior practices and is highly configurable for complex features. Unlike DKT, LKT does not require iterative pretraining and is highly interpretable [47]. These advantages make LKT a good choice for an adaptive learning system user model. While some might argue the BKT is more popular than other methods, this may be a simple artifact of the high publication rates related to earlier development and research on the Carnegie Learning, Inc. systems [13], [48]. It is difficult to find other companies that use BKT, and other large companies use other methods and regression variants (e.g., Duolingo, Inc.) [49].

*C. Goals of the Paper*

At the heart of this project is the specification of a symbolic notation system for alternative logistic regression models that is powerful enough to specify many past models in the literature. In addition to providing a language to compare existing models, the notation system reveals a pallet of features that can be combined in a linear equation to specify novel models. The system allows more precise communication about the differences between 2 models, since the differences can be precisely specified using the models' symbolic notation.

The notation system involves specifying the number of terms, each describing a feature of the data for some factor (we call these components). Components are split into levels, each describing a subset of data for which a feature applies (e.g., student or knowledge component). The most straightforward feature is the intercept, which results in a constant coefficient for each factor's level. Another broad type of feature is the linear feature, where the effect is a linear function of some prior count of some event type within each level of the factor. Linear features are used in the AFM and PFA models [6], [7]. The most complex feature type is a nonlinear feature, where a nonlinear function is applied to data (in contrast to a linear function of counts of practice). Some nonlinear functions also require nonlinear parameters. Fig. 1 shows an example of this notation for each term, where the feature is listed in regular font with the component noted in subscript, such as feature$_{component}$.

$$\ln \frac{p(success)}{1-p(success)} = \text{propdec}_{Student} + \text{intercept}_{Item} + \text{logsuc}_{KC} + \text{logfail}_{KC} + \text{recency}_{Itemx}$$

Fig. 1. Example model.

This paper's primary goal is to demonstrate the utility of LKT by expressing several previously developed and popular models (e.g., AFM, PFA, R-PFA) and new models (e.g., Base4). This will show how LKT is a system that is general to the specific model. The models will be illustrated using several datasets, further validating the method's generality. The models also indicate why no specific model is likely to be "best" for all circumstances by showing differences by dataset. However, some features appear to be generally useful to include. We focus on models that are easily implementable in an adaptive learning system by avoiding models (in the main comparison) with fixed or random intercepts for individual students. While a pretest could be used to estimate student-level model intercept values, we do not address this possibility here. Instead, the models here attempt to avoid student parameters as a partial solution to this cold start problem that occurs when models are transferred between different populations of students.

Because of this focus on models that require no prior training of subject parameters, this paper aims to use a feature (propdec) to trace the subject level component of performance during learning. We will apply a feature from the R-PFA model [3] but repurpose it in modified form to predict subject performance in addition to KC performance. We used this modified R-PFA as a component of our main comparison models after comparing this new component with other possible subject tracking features and fixed and random student intercept models. We demonstrated it captured similar amounts of variance (in Table IV). An in-depth analysis of the options for capturing subject level variance is beyond the scope of this paper.

In addition to our goal of introducing the LKT system for building adaptive learning models, we illustrate the framework's extensibility by comparing some models that use novel features included in LKT. These other new features might apply to KC or item level components. For example, we introduce four new memory-based features, inspired by or directly transferred from the psychology literature [38]–[42]. The LKT framework contains several other features that are not tested here due to the limits of space. These examples are intended to demonstrate the extensibility of the framework.

## II. LKT (Logistic Knowledge Tracing) Framework

To use LKT, one needs to have:
1. Terms of the model, each including a feature that describes change across repetitions of categories at a component level.
2. A sequence of learner event data with subject and correctness columns (at minimum). Usually, this takes the form of a tab-delimited file with columns for student and correctness. See details below in Section E.

*A. Component Level*

The component specifies the subsets of the data for which the feature applies. We might think of the most fundamental component as the subject itself. There are other components as well, such as the items and knowledge components. In the model, each feature's effects for each component sum together to compute the additive effect of multiple features. It is assumed that, except for constant intercepts, a feature is applied for all component levels individually within the data for each subject.

*1) Items*

An item is a specific practice performance opportunity, like providing the English meaning of the Spanish word "gato" or providing the next step in the problem 5x+6=0. Typically, items are split so that each item is assigned a single target KC (see below) since if a single performance maps to many skills, it becomes hard to assign blame for failure results. The item-level component captures the fact that idiosyncratic differences exist for any specific instantiated problem for a topic or concept. Therefore an item-level component intercept is isomorphic with the difficulty parameter in item-response theory. Such difficulties may come from any source, including the context of the problem, numbers used in the item (e.g., for a math problem), vocabulary used in the item (e.g., for a story problem or essay response item), and any other factors that result in the item being difficult for the average student. While we may consider the existence of item by person interactions (e.g., Item A is easy for Sally, but not for John), they are rarely possible to identify ahead of time and so are not used in the models here.

Item components may also be traced using a learning or performance tracing feature. Items are most simply defined as problems with a constant response to a constant stimulus, and people tend to learn constant responses to exact stimulus repetitions very quickly [24]. Often, item-level learning tracing is not used because adaptive systems are built never to repeat items and focus on KC component level learning and performance tracing. Item-level components in LKT to allow researchers to compare with KC-level models, which may help identify possible model weaknesses and lead to model respecification.

*2) Knowledge components (KC)*

Any common learnable factor in a cluster of items that controls that cluster's performance may be described as a knowledge component. Knowledge components are intended to capture transfer between related tasks so that practicing the component in one item is assumed to benefit other items that also share the component. It is conceivable that performance for an item may depend on one or more knowledge components. In the case where multiple knowledge components are present, it is possible to use probability rules to model situations where knowledge of multiple components is necessary for a student to answer a specific item correctly. However, in the work here, we take the standard compensatory approach, in which the sum of the influence of the knowledge components is used to estimate the performance for the item if multiple KCs influence that performance. This compensatory approach is similar to the LLTM model that sums factors to estimate performance.

*3) Overall individual*

The student-level component is used to make a feature be a function of the student's prior performance from all prior data. In the case of a student intercept, the entire student's data are used in estimating a constant value. Unlike the intercept, it should be noted that most features are dynamic in this method, depending on prior data. This paper's only constant features are intercepts representing subjects (initial models) and intercepts representing items (main comparison models). A "dynamic" feature means that its effect in the model potentially changes with each trial for a subject. Most dynamic features start at a value of 0 and change as a function of the student's changing history as time passes in some learning system.

*4) Other components*

The flexibility of LKT means that users are not limited to the standard components above. For example, if students were grouped into 4 clusters, a column of the data could be used for component levels to fit each cluster using a different intercept.

### B. Features

These are the functions for computing the effect of the components' histories for each student (except for the fixed feature, the constant intercept). Some features have a single term like exponential decay (expdecafm, described in 2.C.6), a transform using the sequence of prior trials and a decay parameter. Other features are inherently interactive, such as base2, which scales the logarithmic effect of practice by multiplying by a memory decay effect term. Other terms like base4 and ppe involve the interaction of at least three inputs. Table I summarizes 25 features currently supported in the LearnSphere and the R package, indicating if the feature is adaptive and/or dynamic, as well as explaining the input data required from the knowledge components. Adaptive features change as a function of the prior practices' outcome (typically correct or incorrect). In contrast, dynamic features change independent of prior practice quality (these features are functions of when or how much prior practice occurs, NOT the quality of those performances). These features are individually described in the next section.

TABLE I
SELECTED FEATURES CURRENTLY SUPPORTED BY LKT

| Feature | Pars | Adaptive | Dynamic | Input data needed for component level (e.g., KCs) for each learner |
|---|---|---|---|---|
| **intercept** | 0 | no | no | Intercepts are an exception since they are fit without regard to subject history |
| **lineafm** | 0 | no | yes | Total practice counts |
| **logafm** | 0 | no | yes | Total practice counts |
| **powafm** | 1 | no | yes | Total practice counts |
| **recency** | 1 | no | yes | Age of most recent trial |
| **expdecafm** | 1 | no | yes | Total practice counts |
| **base** | 1 | no | yes | Total practice counts and age of oldest trial |
| **base2** | 2 | no | yes | Total practice counts, age of oldest trial, and intrasession total time |
| **base4** | 4 | no | yes | Total practice counts, age of oldest trial, and intrasession total time |
| **ppe** | 4 | no | yes | Total practice counts and ages of all trials |
| **logsuc** | 0 | yes | yes | Success counts |
| **linesuc** | 0 | yes | yes | Success counts |
| **logfail** | 0 | yes | yes | Failure counts |
| **linefail** | 0 | yes | yes | Failure counts |
| **expdecsuc** | 1 | yes | yes | Success history |
| **expdecfail** | 1 | yes | yes | Failure history |
| **basesuc** | 1 | yes | yes | Success count and age of oldest trial |
| **basefail** | 1 | yes | yes | Failure count and age of oldest trial |
| **base2fail** | 2 | yes | yes | Failure count, age of oldest trial, and intrasession total time |
| **base2suc** | 2 | yes | yes | Success count, age of oldest trial, and intrasession total time |
| **linecomp** | 0 | yes | yes | Success and failure counts |
| **prop** | 0 | yes | yes | Success and failure counts |
| **propdec** | 1 | yes | yes | Success and failure histories |
| **logit** | 0 | yes | yes | Success and failure counts |
| **logitdec** | 1 | yes | yes | Success and failure histories |

In the parameters column, 0 indicates a linear feature, and 1 or more indicates how many additional nonlinear parameters the feature requires for estimation.

## C. Feature Descriptions

### 1) Constant (intercept)

This feature is a simple generalized linear model intercept, computed for a categorical factor (i.e., whatever categories are specified by the component factor).

### 2) Total count (lineafm)

This feature is from the well-known AFM model [7], which predicts performance as a linear function of prior experiences with the KC.

### 3) Log total count (logafm)

This predictor has been used in prior work (e.g., [36]) and implies logarithmically decreasing marginal returns for practice as total prior opportunities increase. For simplicity, we add 1 to the prior trial count to avoid taking the log(0), which is undefined.

### 4) Power-decay for the count (powafm)

This feature models a power-law decrease in the effect of successive opportunities. The model describes smaller or greater diminishing marginal returns by raising the count to a positive power (nonlinear parameter) between 0 and 1. It is a component of the predictive performance equation (PPE) model [38], but for applications not needing forgetting, it may provide a simple, flexible alternative to logafm. Fig. 2 shows how for a power value d = 0.45, the unscaled logit contribution is similar to the natural log for the first 10 trials before increasing faster. The curve for the power value d = 0.60 is steeper than either the natural log or the power value d = 0.45 across all trials. The unscaled logit contribution refers to the feature's effect before considering the effect of the logistic regression coefficient, which is fit either overall (i.e., one coefficient for all levels of a component factor) or individually (i.e., using the $ operator to fit a coefficient for each level of the component factor, see section II-D).

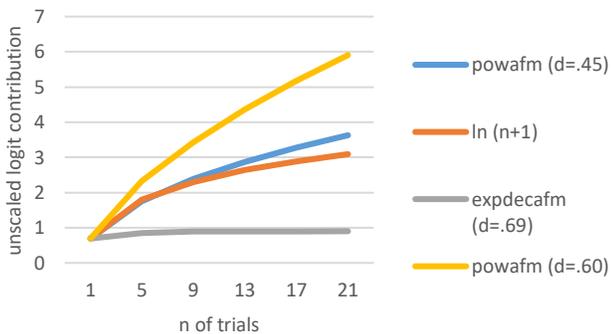

Fig. 2. Graph of logit effect as trials increases with different features.

### 5) Recency (recency)

This feature is inspired by the strong effect of the time interval since the previous encounter with a component (typically an item or KC). This feature was created for this paper and has not been presented previously. This recency effect is well-known in psychology and is captured with a simple power-law decay function to simulate performance improvement when the prior practice was recent. This feature only considers the just prior observation; older trials are not considered in the computation. Fig. 3 shows the basic form of the power-law decay function as a function of time in seconds for different powers. This function assumes time is greater than 1, as this makes the learning equivalent at the start of the function.

### 6) Exponential decay (expdecafm)

This predictor considers the effect of the component as a decaying quantity according to an exponential function. It behaves similarly to logafm or powafm, as shown in Fig. 2. It is similar to the mechanisms first used in the PFA-Decay model [4].

### 7) Power-law decay (base,base2)

This predictor multiplies logafm by the age since the first practice (trace creation) to the power of a decay rate (negative power), as shown in Fig. 3. This predictor characterizes situations where forgetting is expected to occur in the context of accumulating practice effects. Because this factor does not consider the time between individual trials, it implicitly assumes even spacing between repetitions and does not capture recency effects. The base2 version modifies the time by shrinking the time between sessions by some factor, for example, .5, making time between sessions count only 50% towards estimating age. This mechanism to scale forgetting when interference is less was first introduced in cognitive modeling research [40].

### 8) Power-law decay with spacing (base4)

This predictor involves the same configuration as base2, multiplied by the mean spacing to a fractional power. The fractional power scales the effect of spacing such that if the power is 0 or close to 0, then spacing the scaling factor is 1. If the fractional power is between 0 and 1, there are diminishing marginal returns for increasing average spacing between trials. This feature was introduced for this paper as a comparison to the PPE feature.

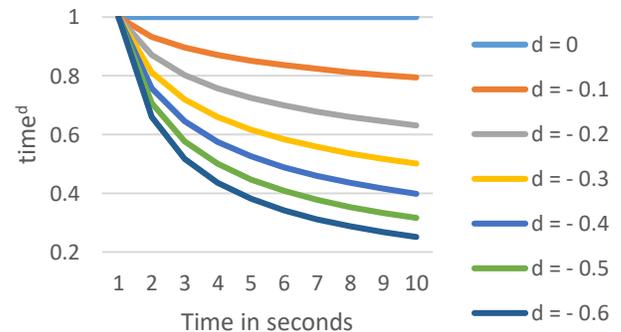

Fig. 3. Graph of age$^{-d}$ effect as age increases with different decay rates.

### 9) Performance Prediction Equation (ppe)

This predictor was introduced over the last several years and shows excellent efficacy in fitting spacing effect data [38], [43]. It is novel because it simultaneously scales practice like the powafm mechanism, captures power-law decay forgetting, spacing effects, and has an interesting mechanism that weights trials according to their recency.

### 10) Log performance measures (logsuc and logfail)

This expression is simply the log-transformed performance factor (total successes or failures), corresponding to the hypothesis that there are declining marginal returns according to a natural log function [8; 26].

*11) Linear PFA (linesuc and linefail)*

These terms are equivalent to the terms in performance factors analysis (PFA) [6], [27], [50], [51].

*12) Exponential decay (expdecsuc and expdecfail)*

This expression uses the decayed count of right or wrong. This method appears to have been first tested by Gong, Beck, and Heffernan [4]. This method is also part of R-PFA, where it is used for tracking failures only, whereas R-PFA uses propdec to track correctness [3]. The function is generally the same as for expdecafm. However, when used with a performance factor, the exponential decay weights on the events seen recently, so a history of recent successes or failures might quickly change predictions since only the recent events count for much, especially if the decay rate is relatively fast.

*13) Linear sum performance (linecomp)*

This term uses the success minus failures to provide a simple summary of overall performance. This feature is parsimonious and less likely to lead to model overfitting or multicollinearity. This feature was created for this paper and has not been presented previously.

*14) Proportion (prop)*

This expression uses the prior probability correct. It is seeded at .5 for the first attempt. This feature was created for this paper and has not been presented previously.

*15) Exponential decay of proportion (propdec and propdec2)*

This expression uses the prior probability correct and was introduced as part of the R-PFA model [3], [52], [53]. This function requires an additional nonlinear parameter to characterize the exponential rate of decay. For propdec, we set the number of ghost successes at 1 and ghost failures at 1 as a modification of Galyardt and Goldin [3]. This modification produces an initial value that can either decrease or increase, unlike the Galyardt and Goldin version (propdec2), which can only increase due to the use of 3 ghost failures and no ghost successes. Our initial comparisons below show that the modified version works to track subject-level variance during learning. Galyardt and Goldin [3] illustrate an extensive number of examples of propdec2's behavior across patterns of successful and unsuccessful trials at various parameter values. The new propdec behaves analogously, except it starts at a value of .5 to represent the different ratio of ghost success to failure at the beginning of practice. As a point of fact, the number of ghost attempts of each type are additional parameters, and we have implemented two settings: 1 ghost success and 1 ghost failure (propdec), or 3 ghost failures (prodec2).

*16) Logit (logit)*

This expression uses the logit (natural log of the success divided by failures). This function requires an additional nonlinear parameter to characterize the initial number of successes or failures. This feature was created for this paper and has not been presented previously.

*17) Exponential decay of logit (logitdec)*

This expression uses the logit (natural log of the success divided by failures). Instead of using the simple counts, it uses the decayed counts like R-PFA, assuming exponential decay and 1 ghost success and 1 ghost failure. This feature was created for this paper and has not been presented previously.

### D. Feature Types

The standard feature type (except for intercept, which is always fit with a coefficient for each level) is fit with the same coefficient for all levels of the component factor. Features may also be fit for each level of the component feature with the $ operator, which causes LKT to use the R model formula operator ":" to fit a coefficient for each level of the component factor (which indicates a regression interaction without main effects in R notation). The most straightforward example of this interaction is for KCs. As shown in Table II, all the models use $ to indicate they fit a different coefficient for each KC. For example, in AFM variants, each KC gets a coefficient to characterize how fast it is learned across opportunities. If a $ operator is not present, a single coefficient would be fit for the feature. Not using the $ operator is particularly relevant when finding a student-level predictor. The $ operator would produce a unique coefficient for each student, indicating the different effects of the feature depending on the student. In this paper, we are specifically focused on models that generalize to a new set of students, so we wish to avoid models that require such individual student parameters. For this reason, we do not use the $ operator to compute the interaction by student (as shown in Table II). Intercept features can be modified with the @ operator, which produces random intercepts using the lmer package instead of the default fixed intercepts. This feature was included primarily to show the initial comparisons of features for modeling student individual differences in Table III.

### E. Learner Data Requirements

The LKT model relies on data being in the DataShop format, but only some columns are needed for the models. The data must consist of practice events for each student, which minimally includes outcome (i.e., correct or incorrect) and optionally includes column(s) identifying times of each practice event, the item id, and any groupings of items into KCs. Additional columns can be used as categorical or quantitative covariate features.

The six datasets used in this paper were chosen to be representative of variability in real learning: two datasets concerned fact-learning (MH, statistics cloze), some were more procedural (tonal learning), and others were a mixture of learning and application of facts and procedures (Assistments, KDD, Andes). This distinction is important because learning and memory theories make distinct predictions for learning and remembering facts versus procedural information depending on the learning context. The MH dataset has not been publicly released, but the other datasets are available on DataShops, with Memphis DataShop [54] housing the cloze data and the other four datasets residing in the CMU DataShop [55].

### F. Making a Model from Scratch

Starting with a dataset and making a model from scratch in the current system requires the user to choose from the available list of features. Beyond the 12 models shown in this paper, thousands of other models may be created that are effective and reasonable in different contexts. Because of this broad generalizability, we conclude with a brief discussion of how to compose a model using LKT.

Most models begin with some representation of student ability. If the model is meant to generalize to unseen students, this will likely begin with the propdec or logitdec features. Either of these features will quickly capture student variability given a reasonably long sequence of practice data. If the model is meant to have a more post hoc, explanatory function, the overall average student ability can be fully captured by student-level intercepts.

Some representation of the initial difficulty of KCs or items will typically be important since few domains contain skills of equal difficulty. Variable initial difficulty can be accounted for with fixed or random effect intercepts. Fixed-effects may be adequate unless the items are truly sampled from some distribution or if the data are sparse. Fixed effects are typically fit for the data's KC level, but fixed effects may also be fit for the items. Since items are frequently nested within KCs, it is possible to create redundant KC models given the item parameters. Fitting more item-level difficulty parameters makes a model more complex but may be acceptable if there is enough data to estimate item difficulties accurately.

After having accounted for differences prior to practice, differences from practice can be considered. Most importantly, there is the effect of learning, and most models include some term that causes an upward slope to performance as practice accumulates. Lineafm is the most basic of these features. The more complex learning features capturing learning may additionally capture forgetting, decreasing marginal return to practice, or spacing effects.

Finally, another useful contribution of LKT is that it enables modelers to evaluate and compare candidate learner models more efficiently. For instance, after the modeler has developed an initial model that cross validates well with dynamic change features, additional adaptive components can be added. It is often useful to think about "splitting" the dynamic features that work well. After fitting a model with lineafm, the modeler could split the feature and instead use linesuc and linefail. Of course, it is possible that simply splitting will not capture the possible richness of alternative models. For example, the fit statistics of these initial models and the modeler's expertise may lead them to hypothesize that the basefail feature may improve fit due to students forgetting the feedback they receive after incorrect responding. The modeler may ultimately conclude that the most robust parsimonious dynamic change model was a combination of linesuc and basefail. In summary, LKT enables easier navigation through the space of possible learner models, with appropriate tools such as cross-validation and model visualizations being used to confirm that the found models are generalizable.

## III. MODELS

We choose the models for our main comparison to show various existing and new logistic models across multiple datasets. The multiplicity of new features we use contrasts past work, which typically only shares one new feature per paper. As discussed in Section I-C, we specifically choose to create models that did not include student parameters, except for models 7–9 provided for reference in Table III and Table IV. For this reason, the models we present are particularly applicable for actual use in adaptive learning systems, rather than as analytic tools to understand the properties of the domain or for other purposes. In Table II the 12 main models are shown. All of the 12 models will use the propdec feature to adapt to students. Models 1 and 2 are versions of the AFM model [7], differing from each other only due to the nonlinear (log) practice function for KCs in model 2. Models 3 and 4 parallel 1 and 2 in that they are propdec student adaptive versions of the PFA model [6], differing only due to the nonlinear (log) practice function for KCs. Model 5 takes the typically better log versions of PFA and adds a new recency term that gives a boost as a function of when the KC was seen last. Model 6 is the PFA Decay model [4], only differing from the original version because it uses propdec rather than a student intercept. Model 7 uses a simplified R-PFA [3] with only the adaptive KC term, while model 8 uses the standard R-PFA terms (but see above regarding propdec vs. propdec2) with the same adaptive student term. Model 9 attempts to improve on R-PFA by making the count of failures sensitive to recency. Models 1012– use the knowledge tracing term from R-PFA (like model 7), but instead of adding failure tracking, memory tracking is added. Model 10 does that with the ppe feature [38], while models 11 and 12 use features new to this paper to track forgetting (model 11) and spacing effects and forgetting (model 12). In short, the new features are motivated by two insights about feature design. The first insight is that for any particular learning effect, such as recency or forgetting, there may be multiple ways to represent it in a model, and some ways will work better to predict learning. The second insight is that we can design new features by turning to cognitive psychology to provide hypotheses about important learning effects that may be captured by new dynamic or adaptive features. Indeed, the original PFA model was based on insight from a long tradition in cognitive psychology, and the new memory and spacing based features are similarly inspired.

These new and recombined models are made possible by how easy it is to add a new feature to the LKT architecture. Features can be combined with other features to create complex hybrid models. We felt it important that the system allowed many options for replication so that prior ideas can be compared with new options. Table II shows the 12 models we compared for each dataset.

TABLE II
MAIN COMPARISON MODELS FOR EACH DATASET

| # | Model Features | Variant of |
|---|---|---|
| 1 | propdec$_{Student}$ + intercept$\$_{KC}$ + lineafm$\$_{KC}$ | AFM |
| 2 | propdec$_{Student}$ + intercept$\$_{KC}$ + logafm$\$_{KC}$ | AFM |
| 3 | propdec$_{Student}$ + intercept$\$_{KC}$ + linesuc$\$_{KC}$ + linefail$\$_{KC}$ | PFA |
| 4 | propdec$_{Student}$ + intercept$\$_{KC}$ + logsuc$\$_{KC}$ + logfail$\$_{KC}$ | PFA |
| 5 | propdec$_{Student}$ + intercept$\$_{KC}$ + logsuc$\$_{KC}$ + logfail$\$_{KC}$ + recency$\$_{KC}$ | PFA |
| 6 | propdec$_{Student}$ + intercept$\$_{KC}$ + expdecsuc$\$_{KC}$ + expdecfail$\$_{KC}$ | PFA-Decay |

| # | Model Feature(s) | |
|---|---|---|
| 7 | propdec$_{Student}$ + intercept\$$_{KC}$ + propdec\$$_{KC}$ | R-PFA |
| 8 | propdec$_{Student}$ + intercept\$$_{KC}$ + propdec\$$_{KC}$ + logfail\$$_{KC}$ | R-PFA |
| 9 | propdec$_{Student}$ + intercept\$$_{KC}$ + propdec\$$_{KC}$ + expdecfail\$$_{KC}$ | new |
| 10 | propdec$_{Student}$ + intercept\$$_{KC}$ + propdec\$$_{KC}$ + ppe\$$_{KC}$ | PPE |
| 11 | propdec$_{Student}$ + intercept\$$_{KC}$ + propdec\$$_{KC}$ + base2\$$_{KC}$ | new |
| 12 | propdec$_{Student}$ + intercept\$$_{KC}$ + propdec\$$_{KC}$ + base4\$$_{KC}$ | new |

There were 100 runs for each model, and we provide the average test values for McFadden's $R^2$ and AUC overall. We also provide the mean test RMSE at the subject level (the same values are used to compute the t-test comparison). Each model was tested using student stratification 100 times using a split-half holdout validation procedure with bootstrapped t-tests. This procedure was designed to allow statistical inference (paired t-tests) where the assumptions of the statistic are fully supported. The 100 runs allowed us to estimate a stable mean t-score difference between each of the models. To do this, we took the data for the held-out fold and computed the error for each held out subject for each model. Using the error for each subject with each model, we could compute the differences in error for the same group of held out subjects for each pairwise model comparison. Since these data are independent, the data are appropriate for a valid paired t-test to see if the difference is greater than 0. We averaged the t-values across the 100 runs to get a more accurate estimate of the test statistic, but this had no influence on power, only accuracy. We include these significance tests after describing the fit of each model.

Since all 12 of these models use the propdec feature at the student level to dynamically capture student variability (as an alternative to random effects or fixed constants), a first analysis focused on 9 initial models to facilitate a comparison of the propdec, propdec2, random and fixed intercepts, and another possible feature to represent students, logitdec (see Table III). Since these models had student intercept parameters, they were not appropriate to cross-validate using our student stratification methods because they cannot be generalized to new students. These comparisons have many fewer parameters than the more complicated main comparison in Table II (which cross-validate without issues, as we can see below), so they can be expected to be stable and were not crossvalidated. We also provide AFM, PFA, and IRT model fits for comparison to the main models.

We applied the same filtering procedure to all six datasets. Students within each dataset were omitted unless they had at least 25 observations. KCs within each dataset were omitted unless they had at least 600 observations. If a student only experienced an example of a KC once, that observation was excluded. Extreme trial duration outliers ( > 95th percentile) were winsorized to equal the 95$^{th}$ percentile trial duration values. Missing trial duration values were imputed with the overall median trial duration. We applied this procedure to all datasets to maintain consistency and interpretability of results while also ensuring reliable parameter estimates. For some datasets, students were provided hints and additional attempts after making errors (Assistments, KDD, Chinese tones, Andes). We chose only to analyze the correctness of students' initial attempts at the first steps of problems. For example, if a student encountered a problem and was initially incorrect but was eventually correct after being provided additional hints, only that initial attempt was included. If the student reencountered the same problem later, again, only the initial attempt would be included. There are practical and theoretical reasons for this focus. For example, if a student answers a multiple-choice problem and is initially incorrect (and told that they are), accuracy will necessarily increase for subsequent attempts immediately following that feedback purely due to the process of elimination (under the assumption that the student would not pick the same incorrect answer). Second, in many cases, the hints are progressively stronger, eventually providing the correct answer. In such situations, the meaning "correctness" has changes, and it may now be merely tracking the students' ability to read the prompt. Finally, our focus is to fit models that track performance and learning over time and across attempts, not the cognitive processes associated with learning from feedback (or following prompts) immediately after the student provides an incorrect answer. Modeling such cognitive processes is a valuable research area but is beyond the scope of the present work.

TABLE III
STUDENT VARIANCE ONLY MODELS USED IN INITIAL COMPARISON FOR EACH DATASET

| # | Model Feature(s) |
|---|---|
| 1 | intercept@$_{Student}$ |
| 2 | intercept$_{Student}$ |
| 3 | propdec$_{Student}$ |
| 4 | propdec2$_{Student}$ |
| 5 | logitdec$_{Student}$ |
| 6 | intercept$_{Student}$ + propdec$_{Student}$ |

## IV. RESULTS

### A. Initial Student Feature Comparison for Datasets

Table IV below shows McFadden's $R^2$ results for some models of student individual differences. This analysis aimed to evaluate how the adaptive methods compared to fixed and random intercepts as approaches to capture individual differences. This analysis shows the similarity of fits for the two intercept-based measures and the three adaptive measures. The 6$^{th}$ model shows how little additional benefit is provided by combining the intercept and the propdec features. This lack of additional benefit suggests that the adaptive propdec feature is highly colinear with the intercept.

TABLE IV
MCFADDEN'S $R^2$ VALUES FOR STUDENT INDIVIDUAL DIFFERENCE MODELS FOR EACH OF THE SIX DATASETS

| # | Model Feature(s) | Cloze | Tone | Assis | KDD | Andes | MH |
|---|---|---|---|---|---|---|---|
| 1 | intercept@$_{Stu.}$ | 0.055 | 0.054 | 0.044 | 0.035 | 0.054 | 0.057 |
| 2 | intercept$_{Stu.}$ | 0.077 | 0.068 | 0.060 | 0.038 | 0.057 | 0.080 |
| 3 | propdec$_{Stu.}$ | 0.070 | 0.066 | 0.062 | 0.057 | 0.076 | 0.060 |
| 4 | propdec2$_{Stu.}$ | 0.077 | 0.054 | 0.066 | 0.054 | 0.074 | 0.066 |
| 5 | logitdec$_{Stu.}$ | 0.071 | 0.066 | 0.061 | 0.057 | 0.070 | 0.060 |
| 6 | intercept@$_{Stu.}$ + propdec$_{Stu.}$ | 0.088 | 0.080 | 0.076 | 0.061 | 0.092 | 0.082 |
| 7 | AFM model | 0.293 | 0.100 | 0.130 | 0.130 | 0.113 | 0.194 |
| 8 | PFA model | 0.300 | 0.104 | 0.132 | 0.139 | 0.122 | 0.204 |
| 9 | 1 param IRT | 0.177 | 0.096 | 0.122 | 0.123 | 0.104 | 0.095 |

All of the models were computed with fixed-effect coefficients (except for #1 shown for comparison purposes).

## B. Statistics Content Cloze Items

The statistics cloze dataset included 58,316 observations from 478 participants who learned statistical concepts by reading sentences and filling in missing words. Participants were adults recruited from Amazon Mechanical Turk. There were 144 KCs in the dataset, derived from 36 sentences, each with 1 of 4 different possible words missing (cloze items). The number of times specific cloze items were presented was manipulated, as well as the temporal spacing between presentations (narrow, medium, or wide). The post-practice test (filling in missing words) could be after 2 minutes, 1 day, or 3 days (manipulated between students). The stimuli type, manipulation of spacing, repetition of KCs and items, and multiple-day delays made this dataset appropriate for evaluating model fit to well-known patterns in human learning data (e.g., substantial forgetting across delays, benefits of spacing). The dataset was downloaded from the Memphis Datashop repository.

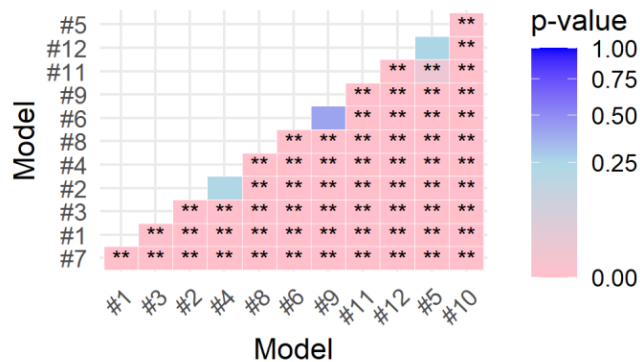

Fig. 4. Statistics cloze p-values, sorted by mean model p-value, for pairwise model comparisons for heldout-subjects. Figure rows and columns are sorted with average lowest p-values on top for the y-axis and the right for the x-axis. * indicates uncorrected p<.05 for the 2-sided test. ** indicates significance after adjusting p-values to control false discovery rate [56].

The results are displayed in Fig. 4 and Table V. The pattern of results indicates that models with features based on theories of declarative memory (models 10–12) significantly improved fit beyond those that were insensitive to memory decay. These memory models' success was consistent with the learning task (participants being asked to recall missing words) and the spacing intervals between practice attempts. However, not all learning may be as vulnerable to decay [57], as suggested by the next dataset.

TABLE V
STATISTICS CLOZE DATA FIT STATISTICS FOR THE 12 MODELS

| # | Model Abbreviation | $R^2_{test}$ | $RMSE_{test}$ | $AUC_{test}$ |
|---|---|---|---|---|
| 1 | lineafm$_{KC}$ | 0.265 | 0.415 | 0.818 |
| 2 | logafm$_{KC}$ | 0.286 | 0.408 | 0.829 |
| 3 | linesuc$_{KC}$ + linefail$_{KC}$ | 0.279 | 0.413 | 0.822 |
| 4 | logsuc$_{KC}$ + logfail$_{KC}$ | 0.294 | 0.408 | 0.83 |
| 5 | logsuc$_{KC}$ + logfail$_{KC}$ + recency$_{KC}$ | 0.342 | 0.392 | 0.855 |
| 6 | expdecsuc$_{KC}$ + expdecfail$_{KC}$ | 0.314 | 0.4 | 0.842 |
| 7 | propdec$_{KC}$ | 0.207 | 0.434 | 0.782 |
| 8 | propdec$_{KC}$ + logfail$_{KC}$ | 0.303 | 0.405 | 0.835 |
| 9 | propdec$_{KC}$ + expdecfail$_{KC}$ | 0.314 | 0.4 | 0.842 |
| 10 | propdec$_{KC}$ + ppe$_{KC}$ | 0.346 | 0.39 | 0.857 |
| 11 | propdec$_{KC}$ + base2$_{KC}$ | 0.339 | 0.393 | 0.853 |
| 12 | propdec$_{KC}$ + base4$_{KC}$ | 0.342 | 0.392 | 0.854 |

## C. Chinese Tone Learning

The Chinese tone learning dataset included 48,443 observations from 97 adult participants enrolled in their first Chinese language course in a US university. Data was collected via an automated tutoring system that provided access to hints after errors (hint requests were treated as incorrect in the following analyses). Only first attempts on the first steps of problems were included for analysis, ultimately retaining 47% of the original dataset. There were 5 KCs, each representing a different tone. These tones differed in their acoustic pitch contours (e.g., rising versus falling pitch). In this study, participants would listen to a recording of a tone and then identify the type of tone by pressing one of five buttons. Depending on which of the three conditions a participant was assigned, the label associated with each button would include either a) a number label (1–5) along with the pinyin (an alphabetic system to aid reading Chinese), b) a visual depiction of the auditory contour of that tone as well as the pinyin, or c) the visual depiction by itself. Only first attempts at classifying the tone were included in the present analysis. Missing values for the trial duration were imputed with the overall median trial duration. The dataset was downloaded from the Carnegie Mellon Datashop repository.

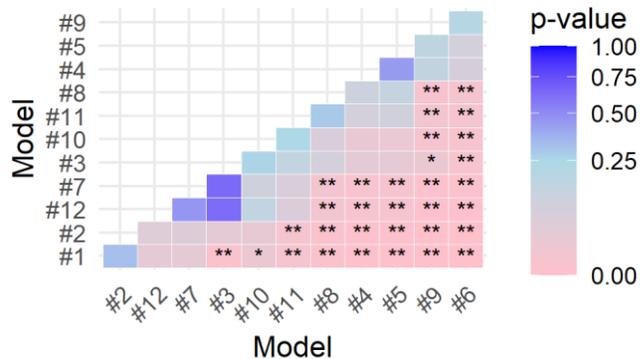

Fig. 5. Chinese tone p-values, sorted by mean model p-value, for pairwise model comparisons for heldout-subjects. Figure rows and columns are sorted with average lowest p-values on top for the y-axis and the right for the x-axis. * indicates uncorrected p<.05 for the 2-sided test. ** indicates significance after adjusting p-values to control false discovery rate [56].

The results are displayed in Fig. 5 and Table VI. In contrast with the cloze dataset, tone learning was not predicted best by the models with memory features. Instead, simpler models that were insensitive to memory decay over time, e.g., Model 8, were sufficient. Differentiating and classifying the perceptual qualities of sounds may not be expected to decay to the same extent as declarative concepts [57]. The distinct differences between the best fitting models for the cloze and tone datasets highlight the need for an overarching framework that allows various knowledge-tracing features to be compared.

TABLE VI
CHINESE TONE LEARNING DATA FIT STATISTICS FOR THE 12 MODELS

| # | Model Abbreviation | $R^2_{test}$ | $RMSE_{test}$ | $AUC_{test}$ |
|---|---|---|---|---|
| 1 | lineafm$_{KC}$ | 0.0984 | 0.385 | 0.698 |
| 2 | logafm$_{KC}$ | 0.0982 | 0.385 | 0.709 |
| 3 | linesuc$_{KC}$ + linefail$_{KC}$ | 0.106 | 0.383 | 0.704 |
| 4 | logsuc$_{KC}$ + logfail$_{KC}$ | 0.111 | 0.381 | 0.722 |
| 5 | logsuc$_{KC}$ + logfail$_{KC}$ + recency$_{KC}$ | 0.113 | 0.381 | 0.724 |
| 6 | expdecsuc$_{KC}$ + expdecfail$_{KC}$ | 0.113 | 0.381 | 0.725 |

| 7 | propdec$_{KC}$ | 0.108 | 0.383 | 0.722 |
| 8 | propdec$_{KC}$ + logfail$_{KC}$ | 0.111 | 0.382 | 0.723 |
| 9 | propdec$_{KC}$ + expdecfail$_{KC}$ | 0.112 | 0.381 | 0.725 |
| 10 | propdec$_{KC}$ + ppe$_{KC}$ | 0.112 | 0.382 | 0.724 |
| 11 | propdec$_{KC}$ + base2$_{KC}$ | 0.111 | 0.382 | 0.724 |
| 12 | propdec$_{KC}$ + base4$_{KC}$ | 0.11 | 0.383 | 0.722 |

### D. Assistments

The Assistments dataset included 580,785 observations from 912 middle school students learning mathematics, collected across 2004/2005. The Assistments tutoring system assists students when they answer questions incorrectly by breaking down the original problem into multiple simpler problems (for a detailed explanation, see [58]). Only first attempts on the first steps of problems were included for analysis, ultimately retaining 23% of the original dataset. The dataset was downloaded from the Carnegie Mellon Datashop repository.

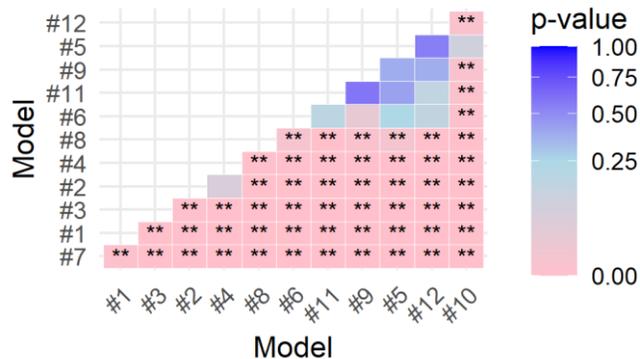

Fig. 6. Assistments p-values, sorted by mean model p-value, for pairwise model comparisons for heldout-subjects. Figure rows and columns are sorted with average lowest p-values on top for the y-axis and the right for the x-axis. * indicates uncorrected p<.05 for the 2-sided test. ** indicates significance after adjusting p-values to control false discovery rate [56].

The results are displayed in Fig. 6 and Table VII. With the Assistments dataset, R-PFA (model 9) and models with memory features (models 10–12) were not significantly different in fit from each other. However, they were generally better than the other models. This pattern suggests that weighting performance according to recency (which all of the better performing models could achieve) was an important factor in this dataset.

TABLE VII
ASSISTMENTS DATA FIT STATISTICS FOR THE 12 MODELS

| # | Model Abbreviation | $R^2_{test}$ | RMSE$_{test}$ | AUC$_{test}$ |
|---|---|---|---|---|
| 1 | lineafm$_{KC}$ | 0.114 | 0.455 | 0.715 |
| 2 | logafm$_{KC}$ | 0.118 | 0.454 | 0.721 |
| 3 | linesuc$_{KC}$ + linefail$_{KC}$ | 0.117 | 0.455 | 0.717 |
| 4 | logsuc$_{KC}$ + logfail$_{KC}$ | 0.12 | 0.453 | 0.721 |
| 5 | logsuc$_{KC}$ + logfail$_{KC}$ + recency$_{KC}$ | 0.125 | 0.452 | 0.726 |
| 6 | expdecsuc$_{KC}$ + expdecfail$_{KC}$ | 0.124 | 0.452 | 0.725 |
| 7 | propdec$_{KC}$ | 0.111 | 0.456 | 0.714 |
| 8 | propdec$_{KC}$ + logfail$_{KC}$ | 0.123 | 0.452 | 0.724 |
| 9 | propdec$_{KC}$ + expdecfail$_{KC}$ | 0.124 | 0.452 | 0.725 |
| 10 | propdec$_{KC}$ + ppe$_{KC}$ | 0.126 | 0.451 | 0.726 |
| 11 | propdec$_{KC}$ + base2$_{KC}$ | 0.125 | 0.452 | 0.725 |
| 12 | propdec$_{KC}$ + base4$_{KC}$ | 0.125 | 0.452 | 0.725 |

### E. KDD Cup

Random subsets of the KDD cup 2005/2006 dataset (809,694 observations) were used in the present analyses. For each of the 100 runs, we randomly choose 120 students of the total 574 students in the file. In the original study, students learned algebra using the Cognitive Tutor system and were given feedback on their responses as well as solution hints (see [59]). Only first attempts on the first steps of problems were included for analysis, which included 87% of observations. There were typically 39 KCs in a run.

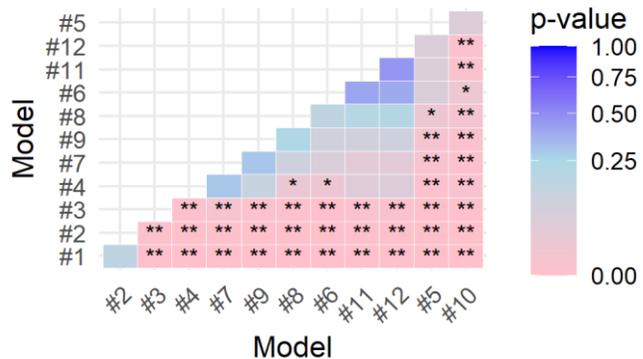

Fig. 7. KDD Cup p-values, sorted by mean model p-value, for pairwise model comparisons for heldout-subjects. Figure rows and columns are sorted with average lowest p-values on top for the y-axis and the right for the x-axis. * indicates uncorrected p<.05 for the 2-sided test. ** indicates significance after adjusting p-values to control false discovery rate [56].

The results are displayed in Fig. 7 and Table VIII. In the KDD cup dataset, models 1–3 tended to provide worse fits than more complex models. As was found with the Assistments dataset, superior models included features that could weight performance according to recency, but explicitly accounting for memory decay was not as relevant.

TABLE VIII
KDD DATA FIT STATISTICS FOR THE 12 MODELS

| # | Model Abbreviation | $R^2_{test}$ | RMSE$_{test}$ | AUC$_{test}$ |
|---|---|---|---|---|
| 1 | lineafm$_{KC}$ | 0.128 | 0.379 | 0.731 |
| 2 | logafm$_{KC}$ | 0.129 | 0.378 | 0.733 |
| 3 | linesuc$_{KC}$ + linefail$_{KC}$ | 0.139 | 0.376 | 0.74 |
| 4 | logsuc$_{KC}$ + logfail$_{KC}$ | 0.147 | 0.374 | 0.749 |
| 5 | logsuc$_{KC}$ + logfail$_{KC}$ + recency$_{KC}$ | 0.161 | 0.372 | 0.761 |
| 6 | expdecsuc$_{KC}$ + expdecfail$_{KC}$ | 0.152 | 0.373 | 0.754 |
| 7 | propdec$_{KC}$ | 0.149 | 0.374 | 0.753 |
| 8 | propdec$_{KC}$ + logfail$_{KC}$ | 0.152 | 0.373 | 0.754 |
| 9 | propdec$_{KC}$ + expdecfail$_{KC}$ | 0.152 | 0.373 | 0.753 |
| 10 | propdec$_{KC}$ + ppe$_{KC}$ | 0.169 | 0.37 | 0.768 |
| 11 | propdec$_{KC}$ + base2$_{KC}$ | 0.154 | 0.373 | 0.755 |
| 12 | propdec$_{KC}$ + base4$_{KC}$ | 0.155 | 0.373 | 0.756 |

### F. Andes

In the Andes dataset, 66 students learned physics using the Andes tutoring system, generating 345,536 observations. Participants were given feedback on their responses as well as solution hints. Additionally, participants were asked qualitative "reflective" questions after feedback (for more details, see [60]). Only first attempts on the first steps of problems were included for analysis, which included 36% of the original dataset. There were 45 KCs.

The results are displayed in Fig. 8 and Table IX. For Andes, several different models provided notable benefits over the simpler models. The PPE model (#10) was again the best fitting model. Models 11–12 (that have some similarities to PPE) also fit well. Interestingly, model 5 was the second-best fitting model. Inspecting the differences in performance between models 5 and 4 indicates that the recency feature may have been
11

beneficial (the only difference between those two models).

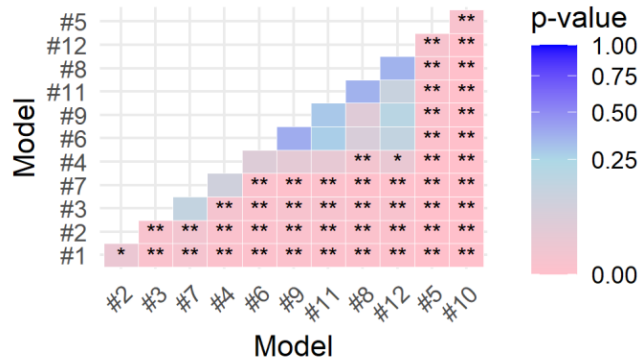

Fig. 8. Andes p-values, sorted by mean model p-value, for pairwise model comparisons for heldout-subjects. Figure rows and columns are sorted with average lowest p-values on top for the y-axis and the right for the x-axis. * indicates uncorrected p<.05 for the 2-sided test. ** indicates significance after adjusting p-values to control false discovery rate [56].

TABLE IX
ANDES DATA FIT STATISTICS FOR THE 12 MODELS

| # | Model Abbreviation | $R^2_{test}$ | $RMSE_{test}$ | $AUC_{test}$ |
|---|---|---|---|---|
| 1 | lineafm$_{KC}$ | 0.131 | 0.394 | 0.732 |
| 2 | logafm$_{KC}$ | 0.131 | 0.394 | 0.733 |
| 3 | linesuc$_{KC}$ + linefail$_{KC}$ | 0.143 | 0.391 | 0.741 |
| 4 | logsuc$_{KC}$ + logfail$_{KC}$ | 0.146 | 0.39 | 0.743 |
| 5 | logsuc$_{KC}$ + logfail$_{KC}$ + recency$_{KC}$ | 0.155 | 0.388 | 0.753 |
| 6 | expdecsuc$_{KC}$ + expdecfail$_{KC}$ | 0.148 | 0.389 | 0.746 |
| 7 | propdec$_{KC}$ | 0.143 | 0.39 | 0.743 |
| 8 | propdec$_{KC}$ + logfail$_{KC}$ | 0.149 | 0.389 | 0.746 |
| 9 | propdec$_{KC}$ + expdecfail$_{KC}$ | 0.148 | 0.389 | 0.746 |
| 10 | propdec$_{KC}$ + ppe$_{KC}$ | 0.165 | 0.386 | 0.762 |
| 11 | propdec$_{KC}$ + base2$_{KC}$ | 0.149 | 0.389 | 0.746 |
| 12 | propdec$_{KC}$ + base4$_{KC}$ | 0.151 | 0.389 | 0.748 |

### G. McGraw Hill

The McGraw Hill dataset contained 124,387 observations from 1047 adult participants. Participants were college students taking coursework on fitness and nutrition. The data was collected from an intelligent tutoring system that accompanied the coursework. Questions in the tutoring system had multiple-choice or multiple-answer formats, and corrective feedback was provided immediately regardless of their correctness. Broadly, the task required recalling information more so than applying skills (in contrast to the Assistments and KDD datasets). There were 111 KCs.

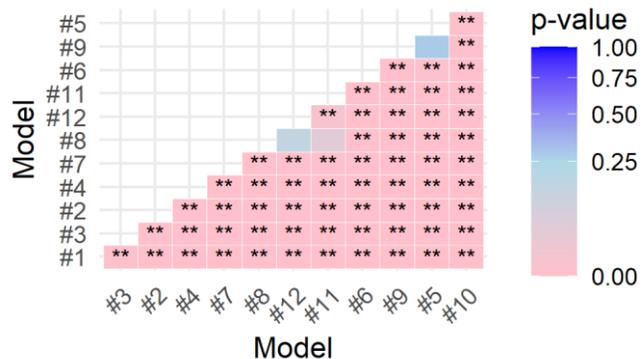

Fig. 9. McGraw Hill p-values, sorted by mean model p-value, for pairwise model comparisons for heldout-subjects. Figure rows and columns are sorted with average lowest p-values on top for the y-axis and the right for the x-axis. * indicates uncorrected p<.05 for the 2-sided test. ** indicates significance after adjusting p-values to control false discovery rate [56].

The results are displayed in Fig. 9 and Table X. Similar to the cloze dataset, for the McGraw Hill dataset, there were large differences in fit between the best model (model 10) and the worst fitting model (model 1). In both cases, students needed to learn and recall factual information, which is more prone to decay over time. Consequently, models that explicitly attempted to track such decay did better (models 10–12). Models that indirectly tracked these changes over time also performed better, such as those using propdec without memory features (models 7–9), exponential decay (model 6), and recency (model 5). Although features like propdec may not have been explicitly developed with memory decay in mind, they can indirectly account for some of the variance tracked by memory-based models.

TABLE X
MCGRAW HILL DATA FIT STATISTICS FOR THE 12 MODELS

| # | Model Abbreviation | $R^2_{test}$ | $RMSE_{test}$ | $AUC_{test}$ |
|---|---|---|---|---|
| 1 | lineafm$_{KC}$ | 0.102 | 0.458 | 0.717 |
| 2 | logafm$_{KC}$ | 0.158 | 0.433 | 0.768 |
| 3 | linesuc$_{KC}$ + linefail$_{KC}$ | 0.109 | 0.457 | 0.718 |
| 4 | logsuc$_{KC}$ + logfail$_{KC}$ | 0.167 | 0.431 | 0.766 |
| 5 | logsuc$_{KC}$ + logfail$_{KC}$ + recency$_{KC}$ | 0.271 | 0.389 | 0.82 |
| 6 | expdecsuc$_{KC}$ + expdecfail$_{KC}$ | 0.268 | 0.391 | 0.819 |
| 7 | propdec$_{KC}$ | 0.165 | 0.423 | 0.769 |
| 8 | propdec$_{KC}$ + logfail$_{KC}$ | 0.217 | 0.408 | 0.795 |
| 9 | propdec$_{KC}$ + expdecfail$_{KC}$ | 0.269 | 0.39 | 0.821 |
| 10 | propdec$_{KC}$ + ppe$_{KC}$ | 0.275 | 0.388 | 0.823 |
| 11 | propdec$_{KC}$ + base2$_{KC}$ | 0.223 | 0.408 | 0.795 |
| 12 | propdec$_{KC}$ + base4$_{KC}$ | 0.22 | 0.408 | 0.794 |

## V. GENERAL DISCUSSION

The results demonstrate the generality of the LKT framework across multiple models and datasets. This paper's analyses demonstrated how LKT could combine various old, modified, and new features to replicate older models and provide examples of new models. The new models outperformed the standard models (e.g., AFM, PFA), demonstrating the utility of LKT. LKT is powerful because these methods are general to a library of features, can be applied broadly to experimental and naturalistic learning data, and allow easy specification of hierarchal models with multiple terms.

LKT shows how logistic regression can serve as an extensible system for learner performance modeling. It is extensible because it is open-source and written in the widely used R statistical programming language. Individual features can be added as part of the LearnSphere project or submitted to our GitHub R repository for inclusion in the main code [61]. Often adding new features requires adding only a few lines of code, after which the new feature can be freely combined to form complex models.

The heart of LKT is the specification system that opens up a new realm of possible models by making explicit the many choices that occur in the process. By making these options explicit, we hope users of the system will arrive at better models through greater awareness of the palette of choices a modeler of learning must make to arrive at a model suitable for their purpose. LKT is intended to make the model-building process more systematic by providing this menu of choices and understanding that ordering off the menu is possible (i.e., adding a new feature to the LKT system).

13*A. Specific Comparisons*

Other than the general utility of the LKT method, there were also some specific results revealed by the comparisons. One interesting result was a contrast in the general model structure for some of the models we tested. Specifically, models 10–12 (the models with memory terms combined with the propdec feature) performed at least as well as the most accurate PFA version, R-PFA. In datasets where students learned more facts instead of procedural skills, models that included memory features tended to outperform those that did not. This advantage has two parts. The first part is that the memory terms in models 10–12 are capturing new variance that was not captured by the expdecfail term in model 9. The second aspect is in the model structure since models 10–12 conveniently separate performance gain into two terms, one term captures the memory benefit that occurs regardless of the success or failure (which assumes feedback is provided in the task), and the other term captures any adjustment to those gains based on tracking the average performance. This model structure may solve one of the key failings of the PFA model methods, which is that they inherently obscure the difference between model adjustments as a function of prior performance and model adjustments as a function of learning. In PFA type models, one term represents the combined learning and performance resulting from prior success, and the other term represents the combined learning and performance resulting from prior failures. This difference may be significant in considerations about which item to choose in an adaptive system since the modeler may wish to select items based on what causes the largest jump in learning rather than the largest jump in performance.

The importance of prior practice's recency as a feature is highlighted by the strength of the R-PFA, PPE, and log PFA with the recency term model (model 5). These three models allow recency effects to dominate the model to the extent they are important for prediction. In R-PFA, proportion correct and count of errors exhibit exponential decay, which makes effects highly dependent on the most recent observations. In PPE, the prior ages of trials can be weighted to recent vs. older, which affects how recency and general forgetting are balanced in the data. In the log PFA model with the simple recency feature, we have further evidence that the recency of similar practices by itself is a highly useful predictor.

The importance of forgetting was highlighted by the PPE, base2, and base4 features in the statistics cloze data results. All three of these models did significantly better than all the recency only models, but only for this dataset. This result was also interesting because this dataset coded the KC at the item level. Hence, repetitions in this model were verbatim repetitions where declarative memory might be expected to be a crucial factor in response accuracy (similar to [41]). In contrast, for the other datasets, base2 and base4 did not provide important advantages. This result suggests that memory features may be useful when a model is used to trace learning for verbatim repetitions of an item.

Other contrasts were similarly interesting and perhaps useful. For example, in all the datasets, the logafm and log PFA models provide large gains relative to the linear models. The consistent advantage of a log transformation indicates that the effect of trials typically shows diminishing marginal returns, a characteristic of the natural log function. These diminishing marginal returns hint at the unexplored possibility that features based on a power function for practice might provide substantial improvement and be more precise in targeting the learning function's curvature. In fact, in addition to the memory factors, PPE uses a power function to scale the effect of prior practice. Thus, PPE combines a flexible practice curve with a similarly flexible forgetting function.

There are also two noticeable patterns in model fit across the different content domains. Specifically, the models were notably worse for the Chinese tones for capturing learning, with only trivial differences in the fit's quality for different models. We suspect this is due to the categorical learning task involved, with only 4 skills for the 4 tones. In this sort of learning, we might not expect much progress with each repetition (slow learning), so the learning terms do not help much. Even when they do, the learning models do not capture transfer between KCs, which may be an important part of category learning (i.e., if a student learns tone 2, tones 1, 3, and 4 all become easier). Similarly, since it is a perceptual task, it is more similar to procedural tasks where forgetting is typically low, so memory factors help little in prediction. Finally, there is strong evidence that the similarity between the present item and recently presented items has important effects on learning when learning categories. For instance, if categories A and B are similar, presenting examples to facilitate comparison across exposures (e.g., ABABAB) may benefit learning more than blocking them (e.g., AAABBB) if the categories are highly similar [62]. This aspect of category learning is distinct from learning in the other datasets and was not directly accounted for with the included model features. Model features could undoubtedly be generated to account for such effects (e.g., [63]). We also expect a model with transfer between the KC would do better for the Chinese tones.

The second pattern is the difference between the memory-based tasks (MHE and Cloze datasets) and the other tasks that are more procedural/conceptual/perceptual. While recency has some benefits in most of the datasets, this benefit and the benefits from general forgetting (which is colinear with recency) in the model are most strong for these two datasets. In both of these datasets, the KCs were individual items, so it was quite sensible that memory factors play a crucial role since the items can be memorized to succeed in the task. For both datasets, the practice's intent was meaningful processing of the verbatim content to produce deep learning, but this does not prevent memory from being a crucial factor in such learning.

*B. Using LKT*

LKT is intended to accelerate future research by making the process of designing adaptive learner models more transparent and systematic. In the past, designers of learner models have moved forward slowly due to the large overhead costs of organizing data and coding the model features with custom-designed code. With LKT, researchers can quickly test many different learner model configurations. In particular, nonlinear features are easy to use and combine in the system, whereas standard logistic regression modeling does not allow for this. The ability to easily fit and compare multiple models may also reveal the subtle behavior of existing features. For instance, although the propdec feature that is part of R-PFA does not track the time between attempts, it does adjust based on recent performance. In many cases, this ability was enough to perform similarly to models that track actual time between attempts.

Interacting with LKT is possible at two levels. First, LKT is implemented in LearnSphere and is immediately available to run on most of the more than 3400 datasets on the DataShop data repository. If the data are in DataShop format, it can be uploaded into the LearnSphere even if it is not already in the DataShop repository, so any data converted to DataShop format can be immediately analyzed. LKT code is also publicly shared as an R package in GitHub [61]. Recent upgrades to both implementations make LKT extremely fast by relying on sparse model matrices and the data.table and LibLineaR packages. These upgrades are allowing analyses previously impossible in base R glm logistic regression. Using the R package devtools, the LKT package can be installed using the command: "install_github("Optimal-Learning-Lab/LKT")".

### C. Limitations and Future Directions

It is important to clarify that this work fits learner models given specific features and an already determined hierarchy (i.e., designations of KCs and students). Another large learner modeling area is concerned with specifying the domain model given the data [7], [64]. In domain model research, the goal is to group the items into KCs, or other often hierarchical groupings, which we assume as input in this paper. Adding such automatic domain model improvement is one future direction we are working on. Our current approach first groups existing KCs or items to form a non-orthogonal Q-matrix where KCs are not independent and then computes an averaged feature as a function of the KCs involved for any observed performance. Recent work has also included other model enhancement features, such as options for testing the model with elastic net fitting, allowing reduction of the model to more compactly represent the set of relevant predictors.

Another limitation is the limited support for random effects, with only intercepts (e.g., random-effect constants) permitted in the syntax. This limitation might be a problem for some theoretical model comparisons. Our primary interests in this paper were evaluating learner models that could be used in an applied context. So we avoided using subject fixed- or random-effects because they are usually difficult to estimate ahead of time. Since the datasets in this paper had many observations for each KC, we choose to fit KC effects as fixed-effect constants since there are typically enough examples for each KC across students to get a reliable fixed estimate. It is also arguable that KCs should be represented as fixed effects because they are independent skills and should not be subjected to random effects shrinkage to a common mean [65]. However, researchers can use random intercepts for KCs when fitting models with LKT.

A next step in this research is to include a latency model to allow users to predict the costs of practice (in units of time) and estimate practice efficiency in terms of a gain per unit of time. With latency predictions, it becomes possible to predict the gain/cost (i.e., the efficiency) for possible practice selections. These predictions can be used to determine practice selection in an adaptive learning system like ACT-R memory models used in prior work [15]. The models presented in this paper are more suitable for mastery determination purposes in their current form, which is a crucial step in the mastery learning algorithm [13], [48]. A goal of the present research was to bring EDM models together with the more complex memory-related features to create a simple and general framework. This paper is the completion of an initial stage in that process.

Finally, after having completed this stage, it has become apparent that the feature specification methods can be enhanced. Currently, interactive features such as base, which multiplies log of the count of prior trials by a forgetting term (see Table I), are fixed in their current formulation. This hardcoding means that making new features that are simple interactions of other features currently requires modifying the underlying R code. A future version of LKT will include enhanced functionality to make interaction terms by composing simple features to create features such as base4 or PPE without having to code them in R.

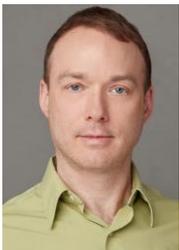

**Philip I. Pavlik Jr.** received his BA degree in economics from the University of Michigan, and Ph.D. in cognitive psychology from Carnegie Mellon University, Pittsburgh, PA, in 2005.

From 2006 to 2011, he was a postdoc and systems scientist at the Carnegie Mellon Human-Computer Interaction Institute and a Pittsburgh Science of Learning Center member. Since 2011 he has been an assistant professor at the University of Memphis, Memphis, TN, promoted to associate professor in 2017. He has published more than 50 articles. His research interests include memory, learning, practice, forgetting, spacing effects, adaptive personalization, mathematical models of learning, and classroom applications of learning technology. He is an Associate Editor for IEEE: Transactions on Learning Technology.

Dr. Pavlik is a current member of the International Society of Educational Data Mining and the Psychonomic Society.

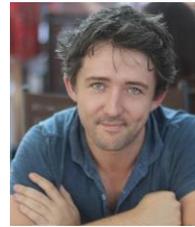

**Luke G. Eglington** received his BS degree in cognitive science from the University of Massachusetts, Boston, MA, in 2013, and a Ph.D. degree in psychological and brain sciences from Dartmouth College, Hanover, NH, in 2018.

Dr. Eglington is currently a Schmidt Futures Learning Engineering postdoctoral fellow at the Institute for Intelligent Systems at the University of Memphis. His research interests include category learning, spacing effects, testing effects, mind-wandering, personalized learning, mathematical models of learning, and interactions between attention and memory.

Dr. Eglington is a current member of the Psychonomic Society.

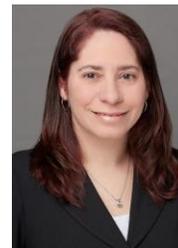

**Leigh M. Harrell-Williams** received her BA degree in Psychology and Business from Virginia Wesleyan College, her MS degree in Statistics from the University of Georgia, Athens, GA in 2004, and her Ph.D. degree in educational research and evaluation at Virginia Tech, Blacksburg, VA, in 2009.

From 2004 to 2012, she was the Undergraduate Coordinate and an Instructor in the Statistics Department at Virginia Tech. From 2012 to 2014, she was a postdoctoral research assistant at Georgia State University. Since 2014, she has been an assistant professor of quantitative research methodology in the Department of Counseling, Educational Psychology, Research, and an affiliate faculty member with the Institute of Intelligent Systems, University of Memphis, Memphis, TN. Her research interests include statistics and mathematics education, social cognitive theory, motivation, and teacher and student beliefs and attitudes.

Dr. Harrell-Williams is an Associate Editor for the Statistics Education Research Journal. She is a member of the American Psychological Association, the American Educational Research Association, the American Statistical Association, and the International Association for Statistics Education.